\documentclass[11pt]{cernrep}
\usepackage{graphicx}
\input{epsf}

\catcode`@=11
\def\citer{\@ifnextchar
[{\@tempswatrue\@citexr}{\@tempswafalse\@citexr[]}}

\def\@citexr[#1]#2{\if@filesw\immediate\write\@auxout{\string\citation{#2}}\fi
  \def\@citea{}\@cite{\@for\@citeb:=#2\do
    {\@citea\def\@citea{--\penalty\@m}\@ifundefined
       {b@\@citeb}{{\bf ?}\@warning
       {Citation `\@citeb' on page \thepage \space undefined}}%
\hbox{\csname b@\@citeb\endcsname}}}{#1}}
\catcode`@=12

\newcommand{\lsim}{\raisebox{-0.13cm}{~\shortstack{$<$ \\[-0.07cm] $\sim$}}~}

\begin{document}
 \title{Higgs Boson Production in Association with Bottom Quarks}
\author{ J. Campbell$^{(a)}$, S. Dawson$^{(b)}$, S. Dittmaier$^{(c)}$,
C. Jackson$^{(d)}$, M. Kr\"amer$^{(e)}$, F. Maltoni$^{(f)}$, L.
Reina$^{(d)}$, M. Spira$^{(g)}$, D. Wackeroth$^{(h)}$, S.
Willenbrock$^{(i)}$}
\institute{
$(a)$ High Energy Physics Division, Argonne National Laboratory, Argonne, Il~~60439, USA\\
$(b)$ Physics Department, Brookhaven National Laboratory, Upton, N.Y.~~11973, USA\\
$(c)$ Max-Planck-Institut f\"ur Physik, F\"ohringer Ring 6, D-80805 M\"unchen, Germany\\
$(d)$ Physics Department, Florida State University, Tallahassee, FL~~32306, USA\\
$(e)$ School of Physics, The University of Edinburgh, Edinburgh~EH9 3JZ, Scotland\\
$(f)$ Centro Studi e Ricerche Enrico Fermi, via Panisperna 89/A, 00184 Rome, Italy\\
$(g)$ Paul Scherrer Institut PSI, CH-5232 Villigen PSI, Switzerland\\
$(h)$ Department of Physics, State University of New York at Buffalo, Buffalo, N.Y.~~14260, USA\\
$(i)$ Department of Physics, University of Illinois at Urbana-Champaign, Urbana, IL~61801, USA}

\maketitle
\begin{abstract}

In the Standard Model, the coupling of the Higgs boson to $b$ quarks
is weak, leading to small cross sections for producing a Higgs boson
in association with $b$ quarks. However, Higgs bosons with enhanced
couplings to $b$ quarks, such as occur in supersymmetric models for
large values of $\tan\beta$, will be copiously produced at both the
Tevatron and the LHC in association with $b$ quarks which will be an
important discovery channel.  We investigate the connections between
the production channels, $b g\rightarrow b h$ and $gg\rightarrow b
{\overline b} h$, at next-to-leading order (NLO) in perturbative QCD
and present results for 
%total cross sections and transverse momentum
%distributions 
the case with two high-$p_T$ $b$ jets and with one
high-$p_T$ $b$ jet at both the Tevatron and the LHC. Finally, the
total cross sections without cuts are compared between $gg\to b\bar b
h$ at NLO and $b\bar b\to h$ at NNLO.

\end{abstract}

\section{Introduction}

In the Standard Model, the production of a Higgs boson in association
with $b$ quarks is suppressed by the small size of the Yukawa
coupling, $g_{bbh}=m_b/v\sim 0.02$. However, in a supersymmetric
theory with a large value of $\tan \beta$, the $b$-quark Yukawa
coupling can be strongly enhanced, and Higgs production in association
with $b$ quarks becomes the dominant production
mechanism.

\begin{figure}[hbt]
\begin{center}
%\vspace*{0.8cm}
\hspace*{-1.5cm} \epsfxsize=7cm \epsfbox{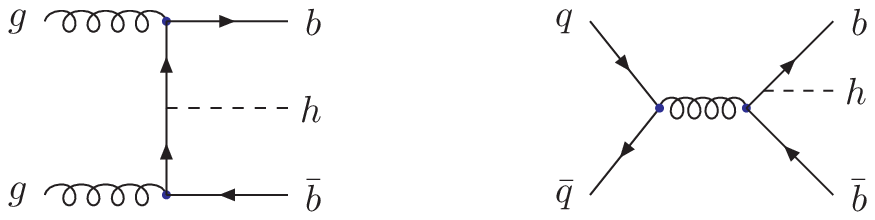}
\caption[ ]{Sample Feynman diagrams for $gg\rightarrow b {\overline b}
h$ and $q {\overline q}\rightarrow b {\overline b} h$ production.}
\label{fg:ggbbh_feyn}
\vspace*{-0.65cm}
\end{center}
\end{figure}

\begin{figure}[hbt]
\begin{center}
%\vspace*{0.8cm}
\hspace*{-1.5cm} \epsfxsize=2.8cm \epsfbox{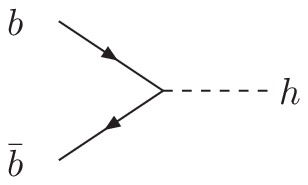}
\caption[ ]{Feynman diagram for $b {\overline b}\rightarrow h$ production. }
\label{fg:bbh_feyn}
\vspace*{-0.65cm}
\end{center}
\end{figure}

\begin{figure}[hbt]
\begin{center}
%\vspace*{0.8cm}
\hspace*{-1.5cm} \epsfxsize=7cm \epsfbox{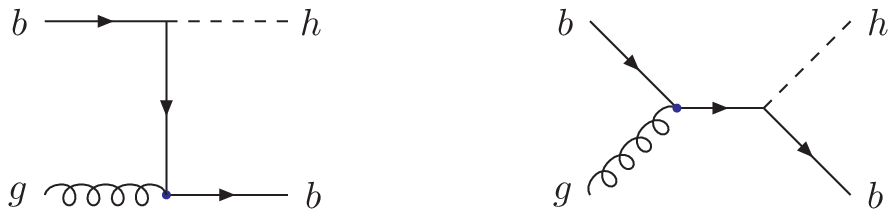}
\caption[ ]{Feynman diagrams for $g b\rightarrow b  h$ production. }
\label{fg:bghb_feyn}
\vspace*{-0.65cm}
\end{center}
\end{figure}

In a four-flavor-number scheme with no $b$ quarks in the initial
state, the lowest order processes are the tree level contributions
$gg\rightarrow b {\overline b} h$ and $q {\overline q}\rightarrow
b {\overline b} h$, illustrated in Fig.~\ref{fg:ggbbh_feyn}.  The
inclusive cross section for $gg\rightarrow b {\overline b} h$
develops potentially large logarithms proportional to $L_b\equiv
\log(Q^2/m_b^2)$ which arise from the splitting of gluons into
$b\bar b$ pairs.\footnote{It should be noted that the $b$ mass in the
argument of the logarithm arises from collinear $b\bar b$
configurations, while the large scale $Q$ stems from $b$ transverse
momenta of this order, up to which factorization is valid.  The scale
$Q$ is the end of the collinear region, which is expected to be of the
order of $M_h/4$~\cite{rsz,msw,bp}.} Since $Q\gg m_b$, the splitting
is intrinsically of ${\cal O}(\alpha_s L_b)$, and because the
logarithm is potentially large, the convergence of the perturbative
expansion may be poor.  The convergence can be improved by summing the
collinear logarithms to all orders in perturbation theory through the
use of $b$ quark parton distributions (the five-flavor-number
scheme)~\cite{dw} at the factorization scale $\mu_F=Q$.  This approach
is based on the approximation that the outgoing $b$ quarks are at
small transverse momentum.  Thus the incoming $b$ partons are given
zero transverse momentum at leading order, and acquire transverse
momentum at higher order.  In the five-flavor-number scheme, the
counting of perturbation theory involves both $\alpha_s$ and $1/L_b$.
In this scheme, the lowest order inclusive process is $b {\overline
b}\rightarrow h$, see Fig.~\ref{fg:bbh_feyn}. The first order
corrections contain the ${\cal O}(\alpha_s)$ corrections to $b
{\overline b}\rightarrow h$ and the tree level process $g b\rightarrow
b h$, see Fig.~\ref{fg:bghb_feyn}, which is suppressed by ${\cal
O}(1/L_b)$ relative to $b {\overline b}\rightarrow h$~\cite{dszw}.  It
is the latter process which imparts transverse momentum to the $b$
quarks.  The relevant production mechanism depends on the final state
being observed.  For inclusive Higgs production it is $b {\overline
b}\rightarrow h$, while if one demands that at least one $b$ quark be
observed at high-$p_T$, the leading partonic process is $gb\rightarrow
b h$.  Finally, if two high-$p_T$ $b$ quarks are required, the leading
subprocess is $g g\rightarrow b {\overline b} h$.

The leading order (LO) predictions for these processes have large
uncertainties due to the strong dependence on the
renormalization/factorization scales and also due to the scheme
dependence of the $b$-quark mass in the Higgs $b$-quark Yukawa
coupling. The scale and scheme dependences are significantly reduced
when higher-order QCD corrections are included.

Section 2 describes the setup for our analysis, and in Section 3 we
compare the LO and NLO QCD results for the production of a Higgs boson
with two high-$p_T$ $b$ jets.  Section 4 contains a discussion of the
production of a Higgs boson plus one high-$p_T$ $b$ jet at NLO,
including a comparison of results within the four-flavor-number and
the five-flavor-number schemes. We consider the corresponding inclusive
Higgs cross sections in Section 5.  Although motivated by the MSSM and
the possibility for enhanced $b$ quark Higgs boson couplings, all
results presented here are for the Standard Model. To a very good
approximation the corresponding MSSM results can be obtained by
rescaling the bottom Yukawa coupling~\cite{dks,djrw}.

\section{Setup}

All results are obtained using the CTEQ6L1 parton distribution
functions (PDFs)~\cite{cteq} for lowest order cross sections and
CTEQ6M PDFs for NLO results.  The top quark is decoupled from the
running of $m_b(\mu)$ and $\alpha_s(\mu)$ and the NLO (LO) cross
sections are evaluated using the $2$~($1$)-loop evolution of
$\alpha_s(\mu)$ with $\alpha_s^{NLO}(M_Z)=0.118$. We use the
${\overline{\rm MS}}$ running $b$ quark mass, $m_b(\mu)$, evaluated at
$2$~($1$)-loop for $\sigma_{NLO}$ ($\sigma_{LO}$), with the $b$ pole
mass taken as $m_b=4.62$~GeV. 
The dependence of the rates on the
renormalization ($\mu_R$) and factorization $(\mu_F)$ scales is
investigated~\cite{dszw,dks,djrw,cemw,hk} in order to
estimate the uncertainty of the predictions for the inclusive Higgs
production channel and for the Higgs plus $1~b$-jet channel.
The dependence of the Higgs plus $2~b$- jet rates on the
renormalization ($\mu_R$) and factorization $(\mu_F)$ scales has been
investigated elsewhere~\cite{dks,djrw} and here we fix
$\mu=\mu_R=\mu_F=(2m_b+M_h)/4$, motivated by the studies in
Refs.~\cite{rsz,msw,bp,dszw,dks,djrw,cemw,hk}.

In order to reproduce the experimental cuts as closely as possible for
the case of Higgs plus 1 or 2 high-$p_T$ $b$ quarks, we require the
final state $b$ and ${\overline b}$ to have a pseudorapidity $\mid
\eta \mid < 2$ for the Tevatron and $\mid \eta \mid < 2.5$ for the LHC.
To better simulate the detector response, the gluon and the
$b/{\overline b}$ quarks are treated as distinct particles only if the
separation in the azimuthal angle-pseudorapidity plane is $\Delta
R>0.4$. For smaller values of $\Delta R$, the four-momentum vectors of
the two particles are combined into an effective $b$/${\overline b}$
quark momentum four-vector.  All results presented in the
four-flavor-number scheme have been obtained independently by two
groups with good agreement~\cite{dks,djrw,dks2,djrw2}.

\section{Higgs + 2 \boldmath{$b$} Jet Production}

Requiring two high-$p_T$ bottom quarks in the final state reduces
the signal cross section with respect to that of the zero and one
$b$-tag cases, but it also greatly reduces the background.  It
also ensures that the detected Higgs boson has been radiated off a
$b$ or ${\overline b}$ quark and the corresponding cross section
is therefore unambiguously proportional to the square of the
$b$-quark Yukawa coupling at leading order, while at
next-to-leading order this property is mildly violated by closed
top-quark loops~\cite{dks,djrw}. The parton level processes
relevant at lowest order are $gg\rightarrow b {\overline b} h$ and
$q {\overline q}\rightarrow b {\overline b} h$, as illustrated in
Fig.~\ref{fg:ggbbh_feyn}.  Searches for the neutral MSSM Higgs
bosons $h,H,A$ produced in association with $b$ quarks have been
performed at the Tevatron~\cite{cdf}.

\begin{figure}[htb]
\begin{center}
\vspace*{5mm}
\includegraphics[bb=50 250 580 600,scale=0.4]{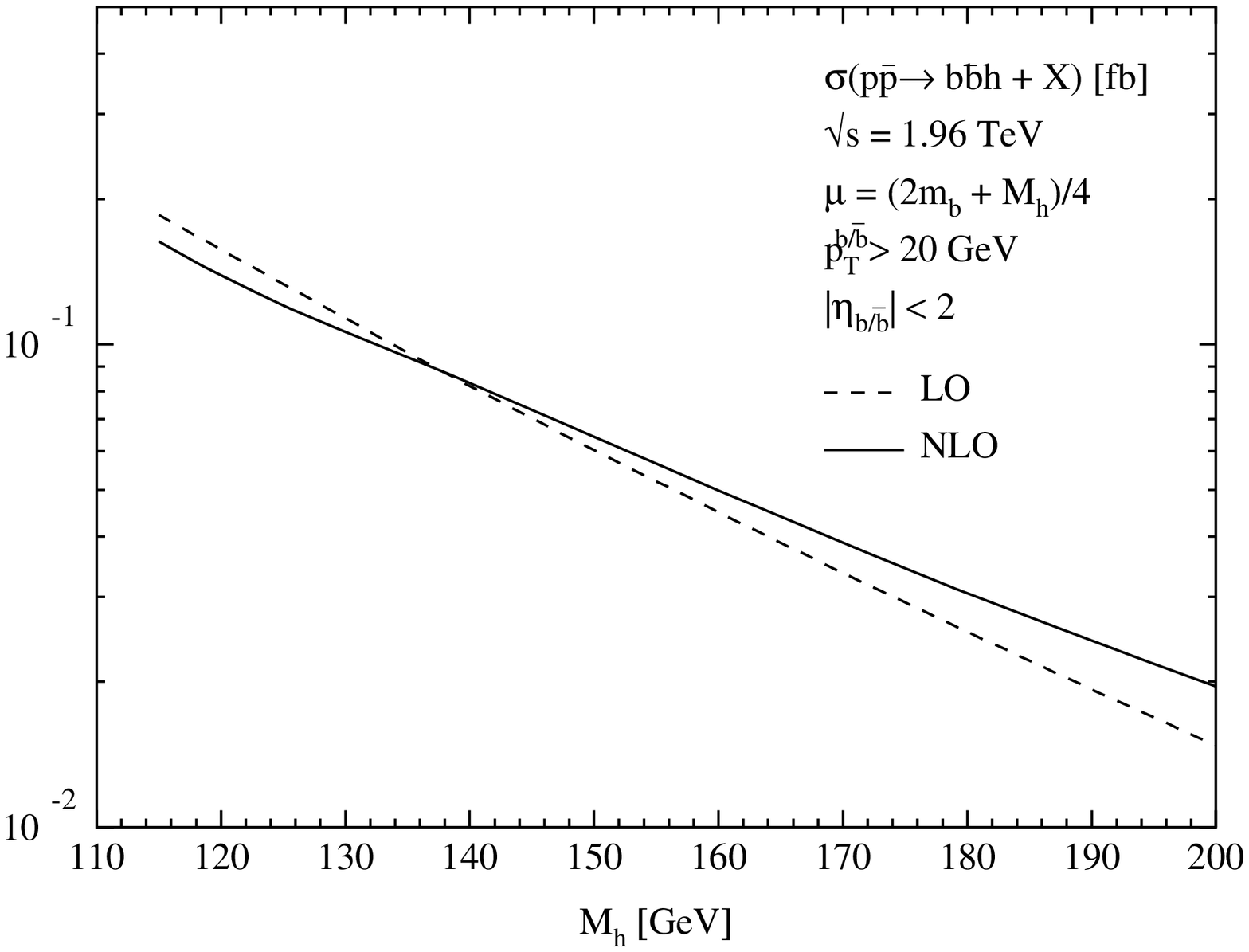}
\includegraphics[bb=50 250 580 600,scale=0.4]{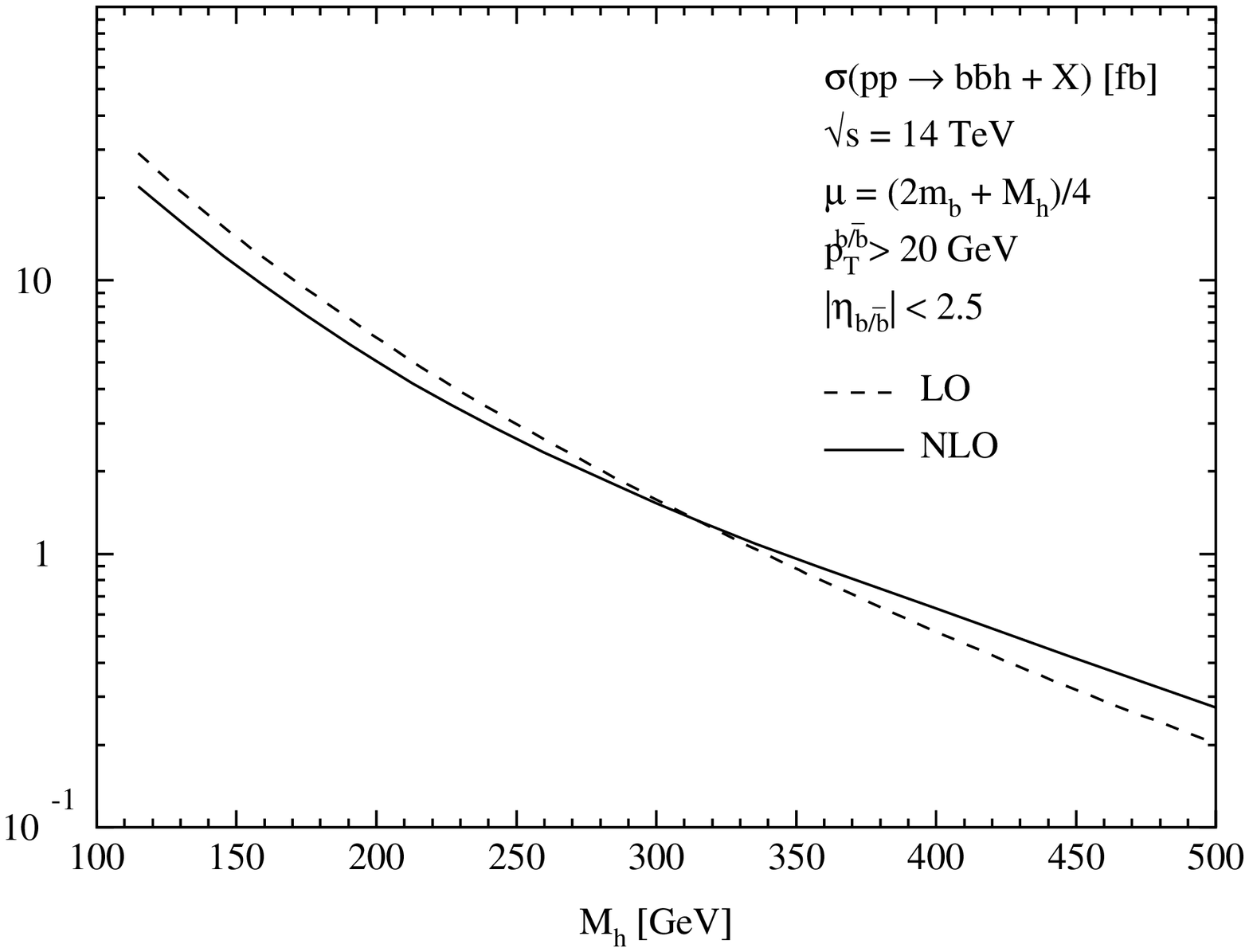}
\caption[]{Total cross sections for $p{\overline p} (pp) \rightarrow b
{\overline b} h+X$ at the Tevatron and the LHC as a function of the
Higgs mass $M_h$ with two high-$p_T$ $b$ jets identified in the final
state.  The $b/\bar b$ quarks are required to satisfy $p_T^{b/\bar
b}>20~GeV$.  We fix $\mu=\mu_R=\mu_F=(2m_b+M_h)/4$.}
\label{fg:2b_sigtot}
\end{center}
\end{figure}

\begin{figure}[htb]
\begin{center}
\includegraphics[bb=50 250 580 600,scale=0.4]{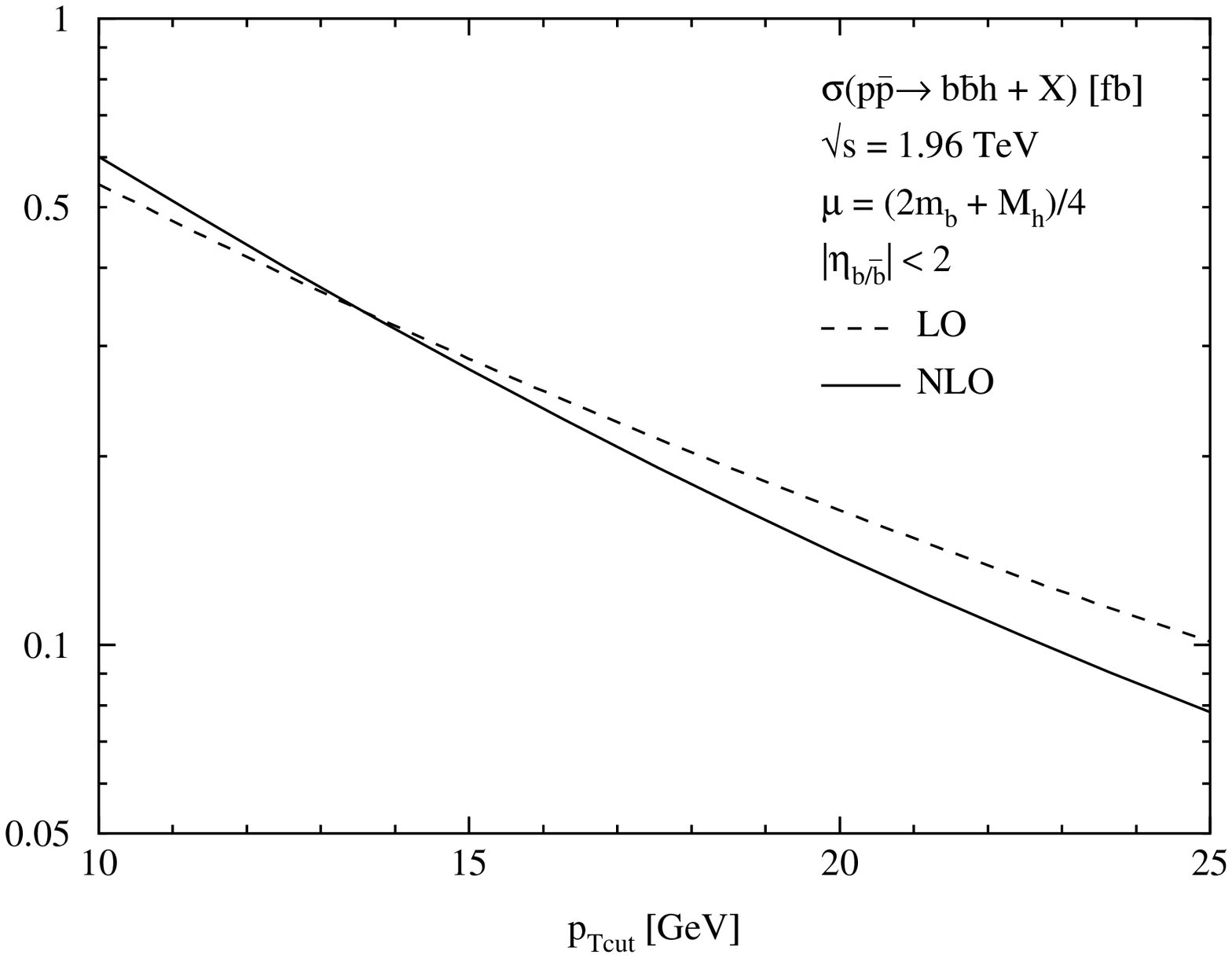}
\includegraphics[bb=50 250 580 600,scale=0.4]{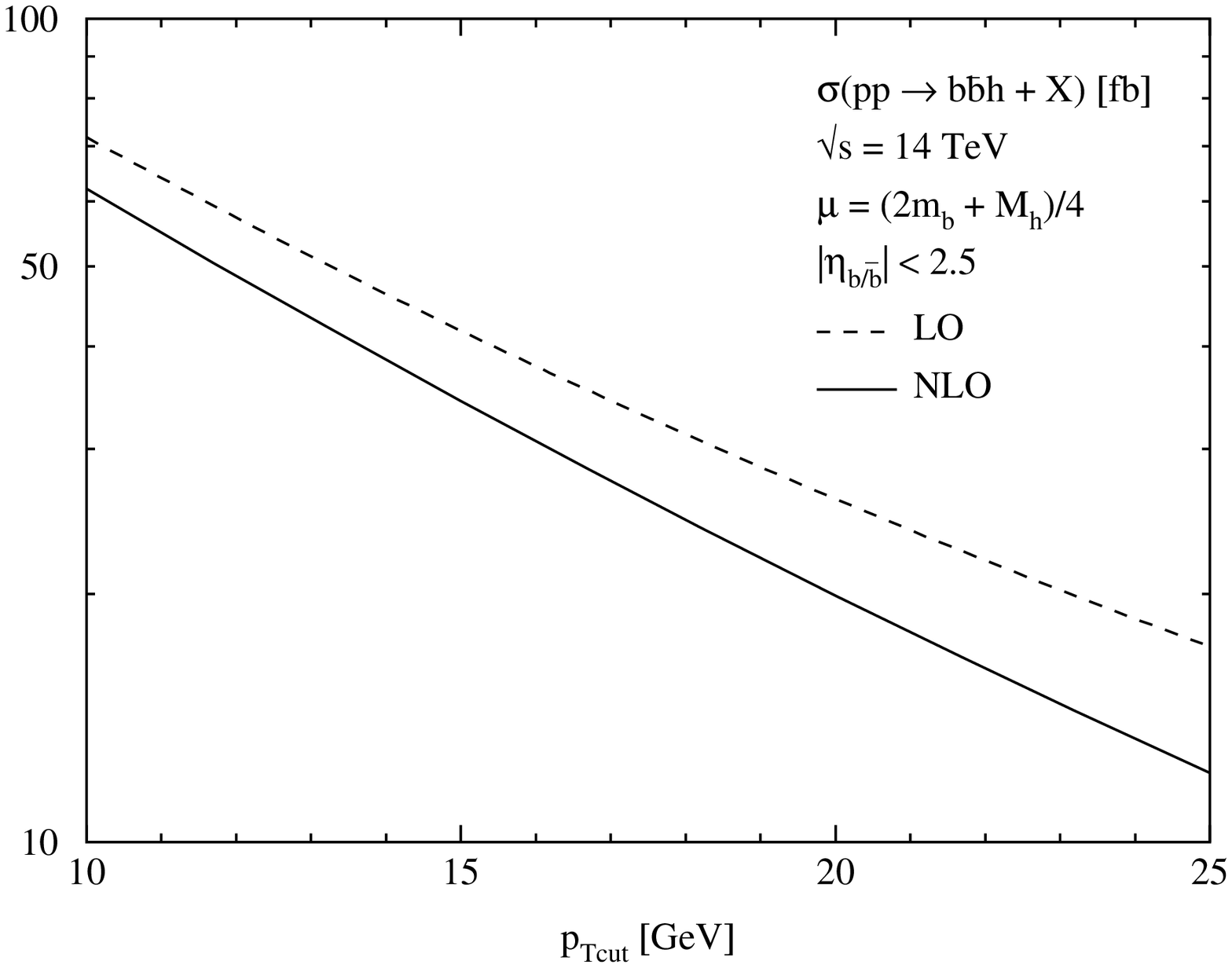}
\caption[]{Total cross sections for $p{\overline p} (pp) \rightarrow b
{\overline b} h+X$ at the Tevatron and the LHC as a function of the
cut $p_{T {\rm cut}}$ in $p_T^{b/\bar b}$ for a Higgs mass $M_h=120$
GeV with two high-$p_T$ $b$ jets identified in the final state.  We
fix $\mu=\mu_R=\mu_F=(2m_b+M_h)/4$.}
\label{fg:2b_ptcut}
\end{center}
\end{figure}

The rate for Higgs plus 2 high-$p_T$ $b$ jets has been computed at NLO
QCD in Refs.~\cite{dks,djrw} and is shown in Fig.~\ref{fg:2b_sigtot}
for both the Tevatron and the LHC. The NLO QCD corrections modify the
LO predictions by $\lsim 30\%$ at the Tevatron and $\lsim 50\%$ at the
LHC. The total cross section plots include a cut on $p_T^{b/{\overline
b}} > 20$ GeV, which has a significant effect on the cross sections.
We show the dependence of the cross section on this cut in
Fig.~\ref{fg:2b_ptcut}. The NLO corrections are negative at large
values of the cut on $p^{b/\bar b}_T$ and tend to be positive at small
values of $p^{b/\bar b}_T$.


\section{Higgs + 1 \boldmath{$b$} Jet Production}

The associated production of a Higgs boson plus a single $b$ quark (or
$\bar b$ quark) is a promising channel for Higgs production in models
with enhanced $b {\overline b} h$ couplings.  The cross section is an
order of magnitude larger than that for Higgs plus 2 high-$p_T$ $b$
jet production for the cuts imposed in our analysis.

\begin{figure}[hbt]
\begin{center}
\vspace*{5mm}
\includegraphics[bb=50 250 580 600,scale=0.4]{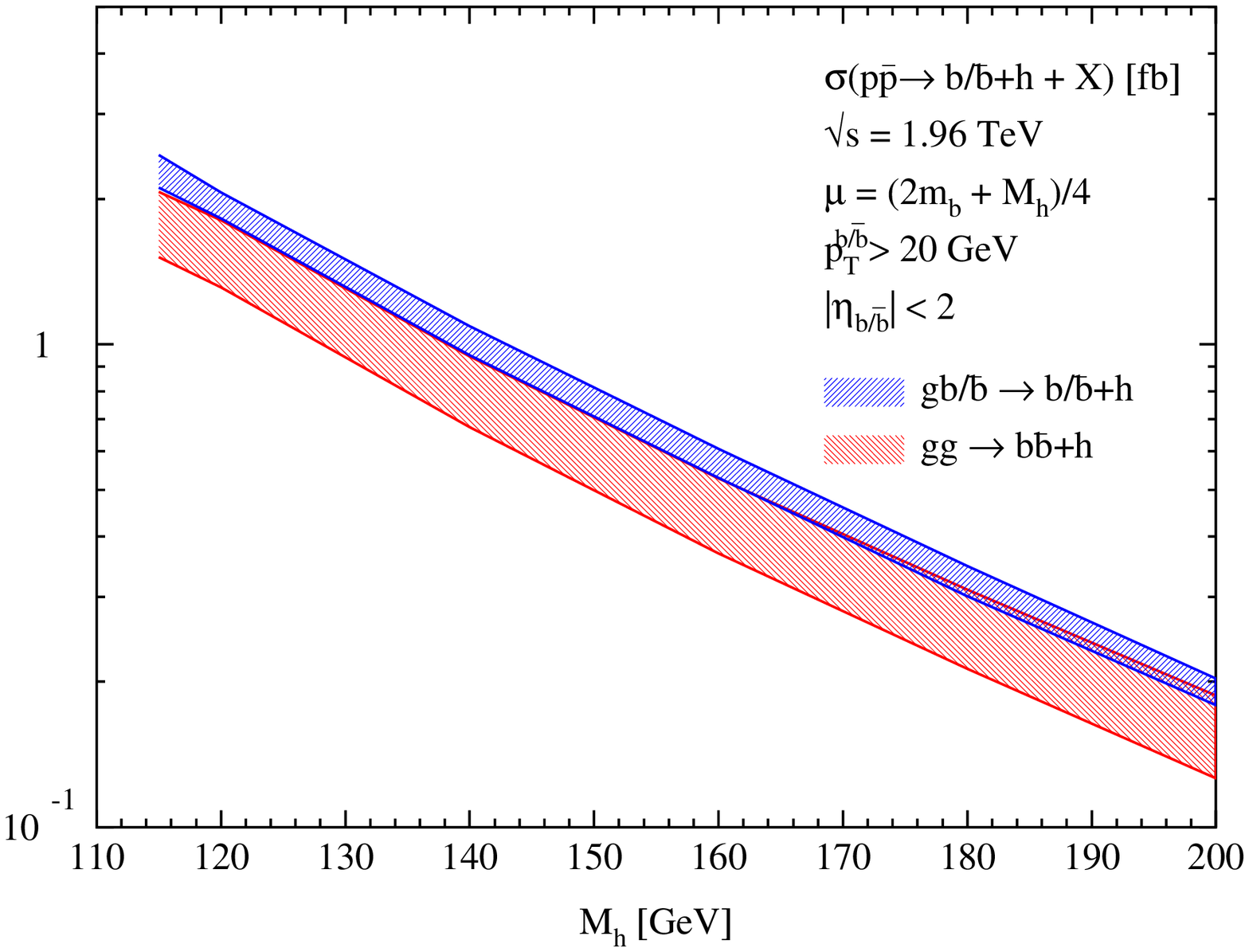}
\includegraphics[bb=50 250 580 600,scale=0.4]{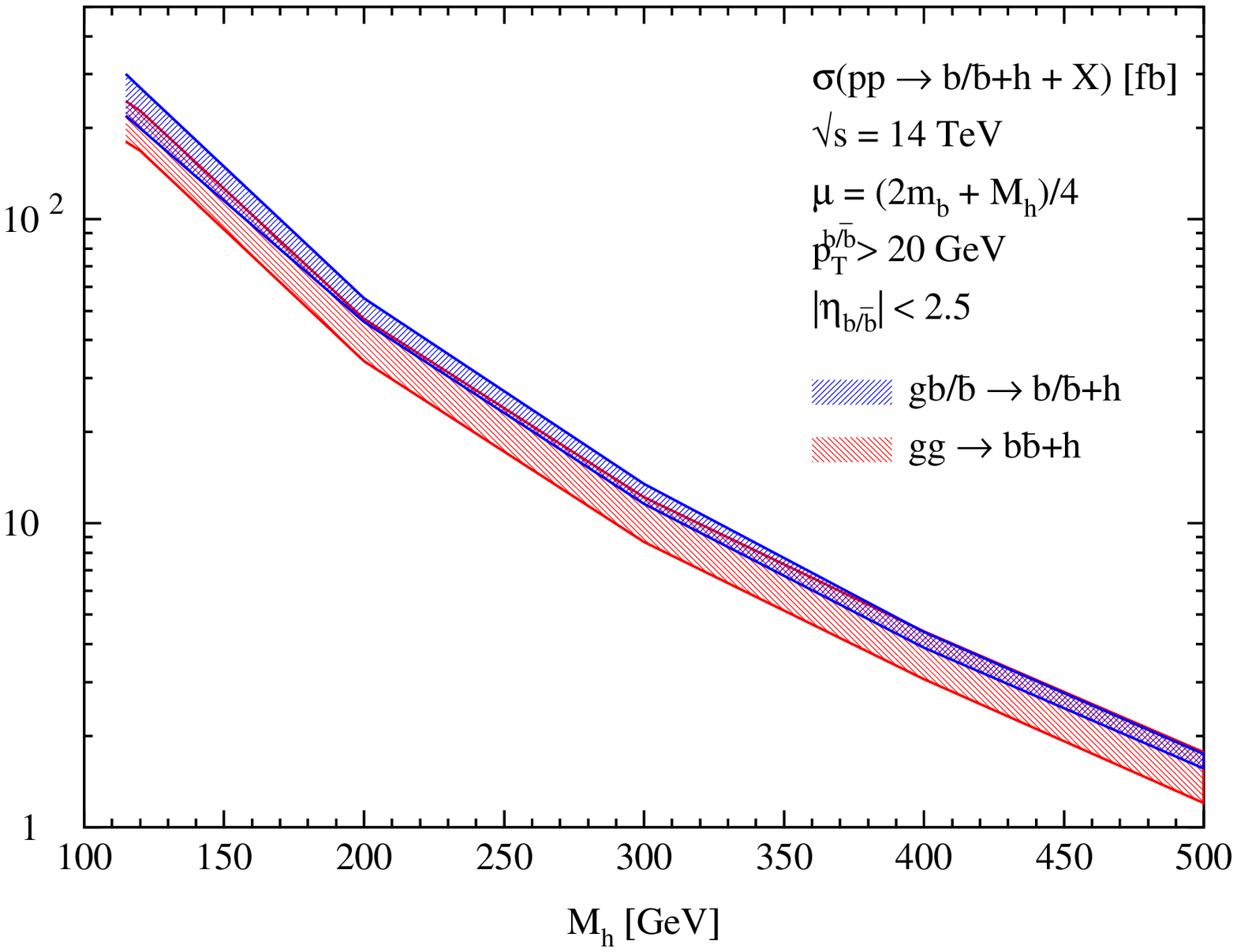}
\caption[]{Total cross sections for $p{\overline p} (pp) \rightarrow b
{\overline b} h+X$ at the Tevatron and the LHC as a function of the
Higgs mass $M_h$ with one high-$p_T$ $b$ jet identified in the final
state.  The $b(\bar b)$ quark is required to satisfy $p_T^{b/\bar
b}>20~\mbox{GeV}$.  We vary the renormalization/factorization
scales around the central 
value $\mu=\mu_R=\mu_F=(2m_b+M_h)/4$ as described in the text.}
\label{fg:1b_sigma}
\end{center}
\end{figure}

\begin{figure}[hbt]
\begin{center}
\includegraphics[bb=50 250 580 600,scale=0.4]{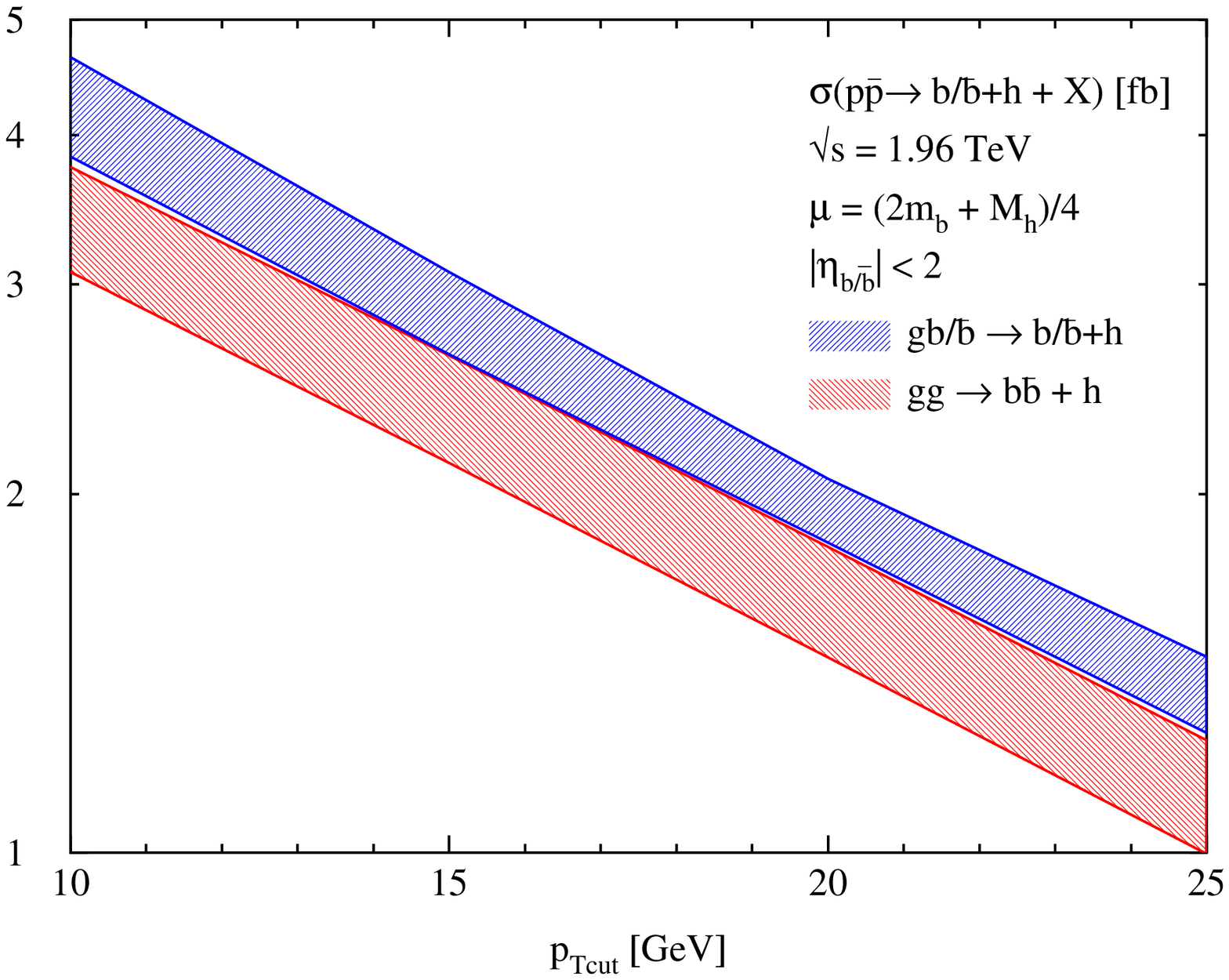}
\includegraphics[bb=50 250 580 600,scale=0.4]{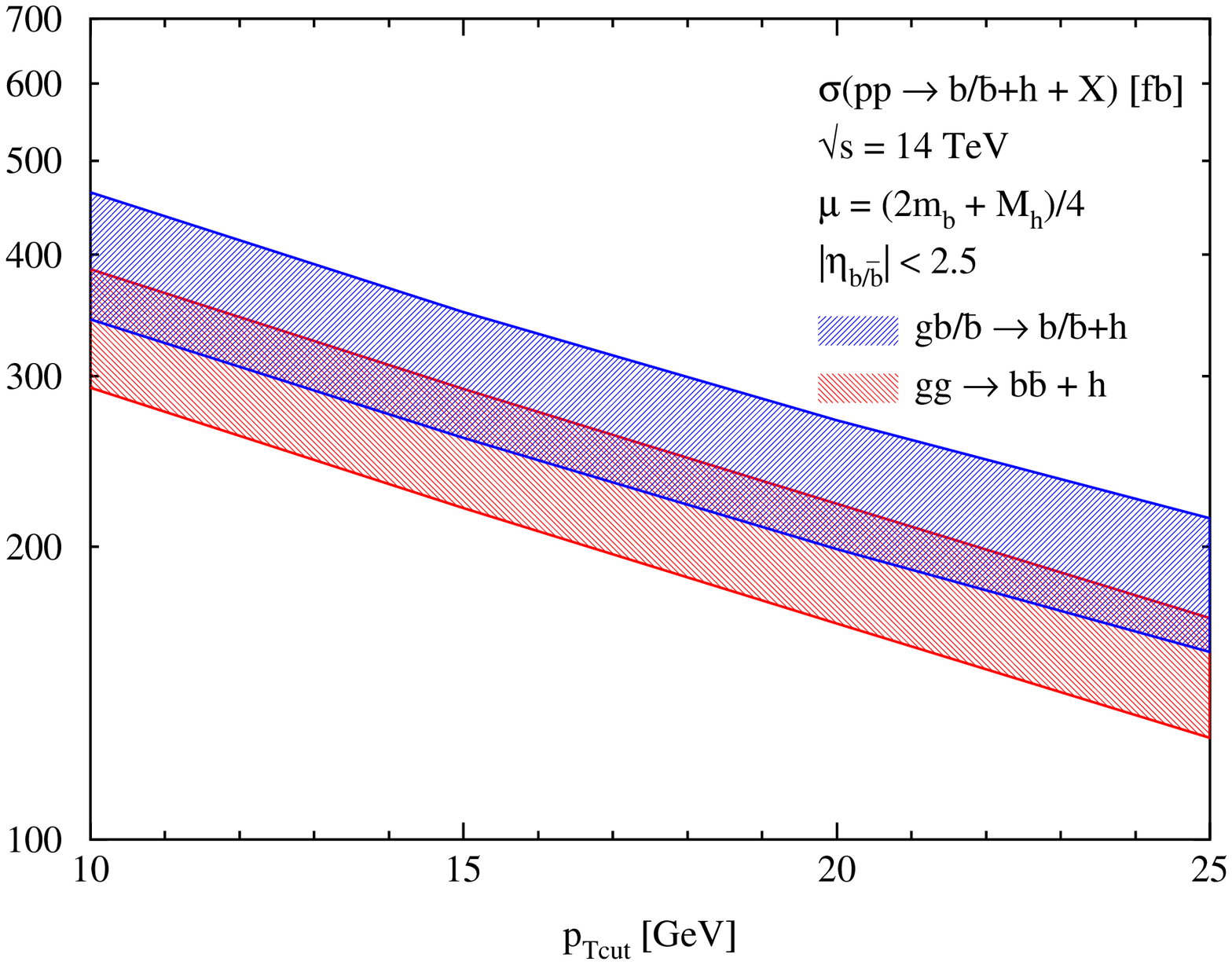}
\caption[]{Total cross sections for $p{\overline p} (pp) \rightarrow b
{\overline b} h+X$ at the Tevatron and the LHC as a function of the cut
$p_{T {\rm cut}}$ in $p_T^{b/\bar b}$ for a Higgs mass $M_h=120$ GeV with one
high-$p_T$ $b$ jet identified in the final state.  
We vary the renormalization/factorization scales around the central 
value $\mu=\mu_R=\mu_F=(2m_b+M_h)/4$ as described in the text.}
\label{fg:1b_ptcut}
\end{center}
\end{figure}

In the four-flavor-number scheme, this process has been computed to
NLO, with the momentum of one of the $b$ quarks integrated
over~\cite{dks,dks2,djrw2}.  This integration yields a potentially
large factor $L_b$. Both the total cross sections and the dependence
on the $p_T^{b,{\overline b}}$ cut at the Tevatron and the LHC are
shown in Figs.~\ref{fg:1b_sigma} and \ref{fg:1b_ptcut}. The NLO
corrections increase the cross section by $\lsim 50\%$ at the Tevatron
and $\lsim 80\%$ at the LHC.
The renormalization/factorization scales are
varied around the central 
value $\mu=\mu_R=\mu_F\equiv (2m_b+M_h)/4$.  
At the Tevatron, the upper bands of
the curves for the four-flavor-number scheme
 in Figs.~\ref{fg:1b_sigma} and
\ref{fg:1b_ptcut} correspond to $\mu_R=\mu_F=2\mu$, while
the lower bands correspond to $\mu_R=\mu_F=\mu/2$.
  The scale dependence is more interesting at the
LHC, where the upper bands are obtained with
$\mu_R=\mu/2$ and $\mu_F=2\mu$, while
the lower bands correspond to $\mu_R=2 \mu$ and $\mu_F=\mu/2$.
At both the Tevatron and the LHC, the width of the error band below
the central value ($\mu=\mu_R=\mu_F$) is larger than above.

In the five-flavor-number scheme, the NLO result consists of the
lowest order process, $b g\rightarrow b h$, along with the ${\cal
O}(\alpha_s)$ and ${\cal O}(1/L_b)$ corrections, which are of moderate
size for our scale choices~\cite{cemw}. The potentially large
logarithms $L_b$ arising in the four-flavor-number scheme have been
summed to all orders in perturbation theory by the use of $b$ quark
PDFs.  
In the five-flavor-number scheme, the upper bands of
the curves for the Tevatron  in Figs.~\ref{fg:1b_sigma} and
\ref{fg:1b_ptcut} correspond to $\mu_R=\mu$ and $\mu_F=2\mu$, while
the lower bands correspond to $\mu_R=\mu/2$ and $\mu_F=\mu$.
At the
LHC, the upper bands are obtained with
$\mu_R=\mu$ and $\mu_F=2\mu$, while
the lower bands correspond to $\mu_R=2 \mu$ and $\mu_F=\mu/2$.
 The two approaches agree within their scale uncertainties, but
the five-flavor-number scheme tends to yield larger cross sections as
can be inferred from Figs.~\ref{fg:1b_sigma} and
\ref{fg:1b_ptcut}.

Contributions involving closed top-quark loops
have not been included in the five-flavor-number scheme
calculation of Ref.~\cite{cemw}.  This contribution is negligible
in the MSSM
for large $\tan\beta$.  In the four-flavor scheme, the closed top-quark
loops have been included and in the Standard Model
 reduce the total cross section for the production of a
Higgs boson  plus a single $b$ jet by 
$\sim -7\%$ at the Tevatron and $\sim -13~\%$ at 
the LHC for $M_h=120$~GeV \cite{dks2,djrw2}.  


\section{Inclusive Higgs Boson Production}

If the outgoing $b$ quarks are not observed, then the dominant process
for Higgs production in the five-flavor-number scheme at large values
of $\tan\beta$ is $b {\overline b} \rightarrow h$.  This final state
contains two spectator $b$ quarks (from the gluon splittings) which
tend to be at low transverse momentum.  At the LHC this state can be
identified through the decays into $\mu^+\mu^-$ and $\tau^+\tau^-$ for
the heavy Higgs bosons $H,A$ at large values of $\tan\beta$ in the
MSSM~\cite{atlascms}.  The $b {\overline b}\rightarrow h$ process has
been computed to NLO~\cite{dszw} and NNLO~\cite{hk} in perturbative
QCD.  The rate depends on the choice of renormalization/factorization
scale $\mu_{R/F}$, and at NLO a significant scale dependence remains.
The scale dependence becomes insignificant at NNLO.  It has been
argued that the appropriate factorization scale choice is
$\mu_F=(M_h+2m_b)/4$~\cite{msw,bp} and it is interesting to note that at
this scale, the NLO and NNLO results nearly coincide~\cite{hk}.

\begin{figure}[htb]
\begin{center}
\vspace*{5mm}
\includegraphics[bb=50 250 580 600,scale=0.4]{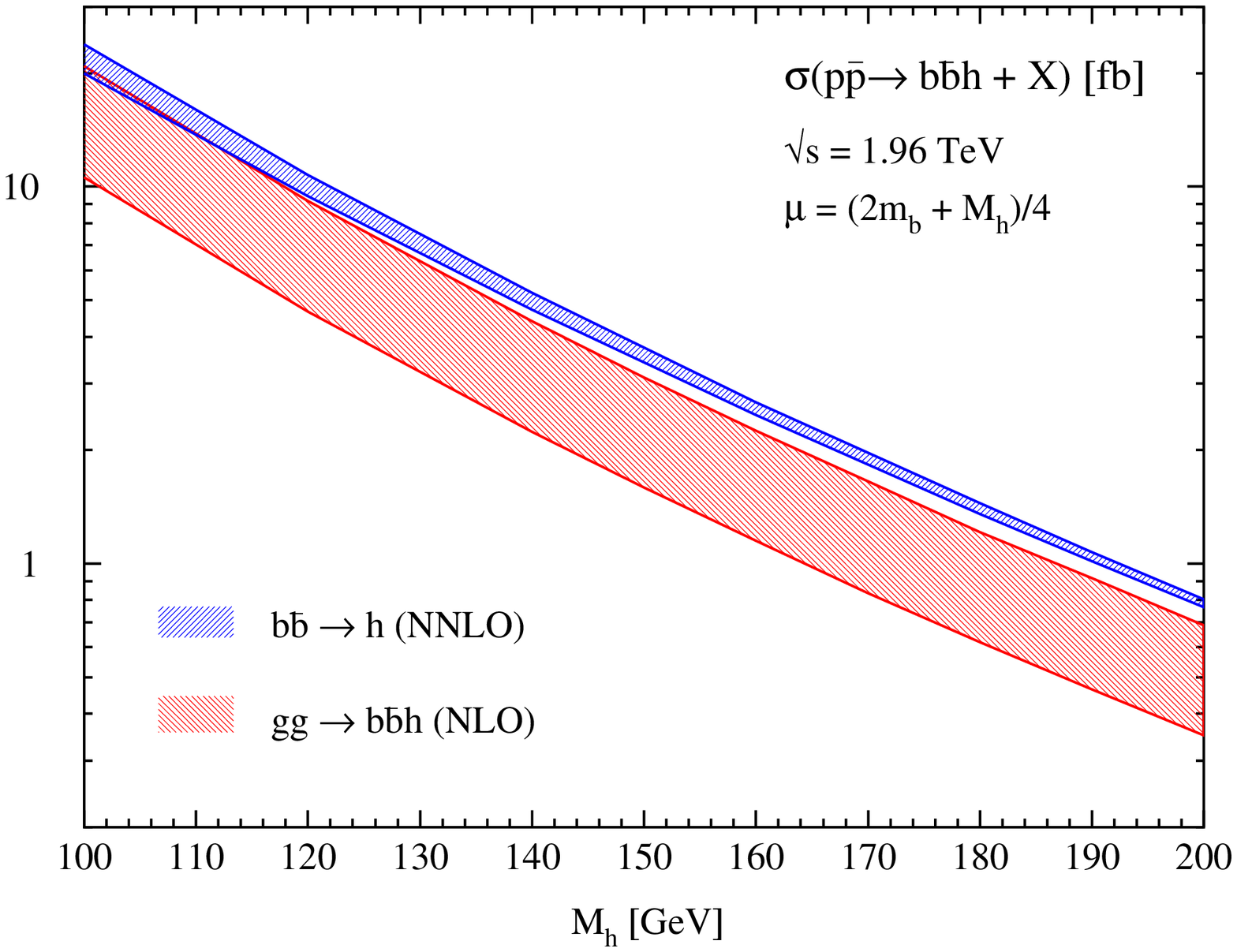}
\includegraphics[bb=50 250 580 600,scale=0.4]{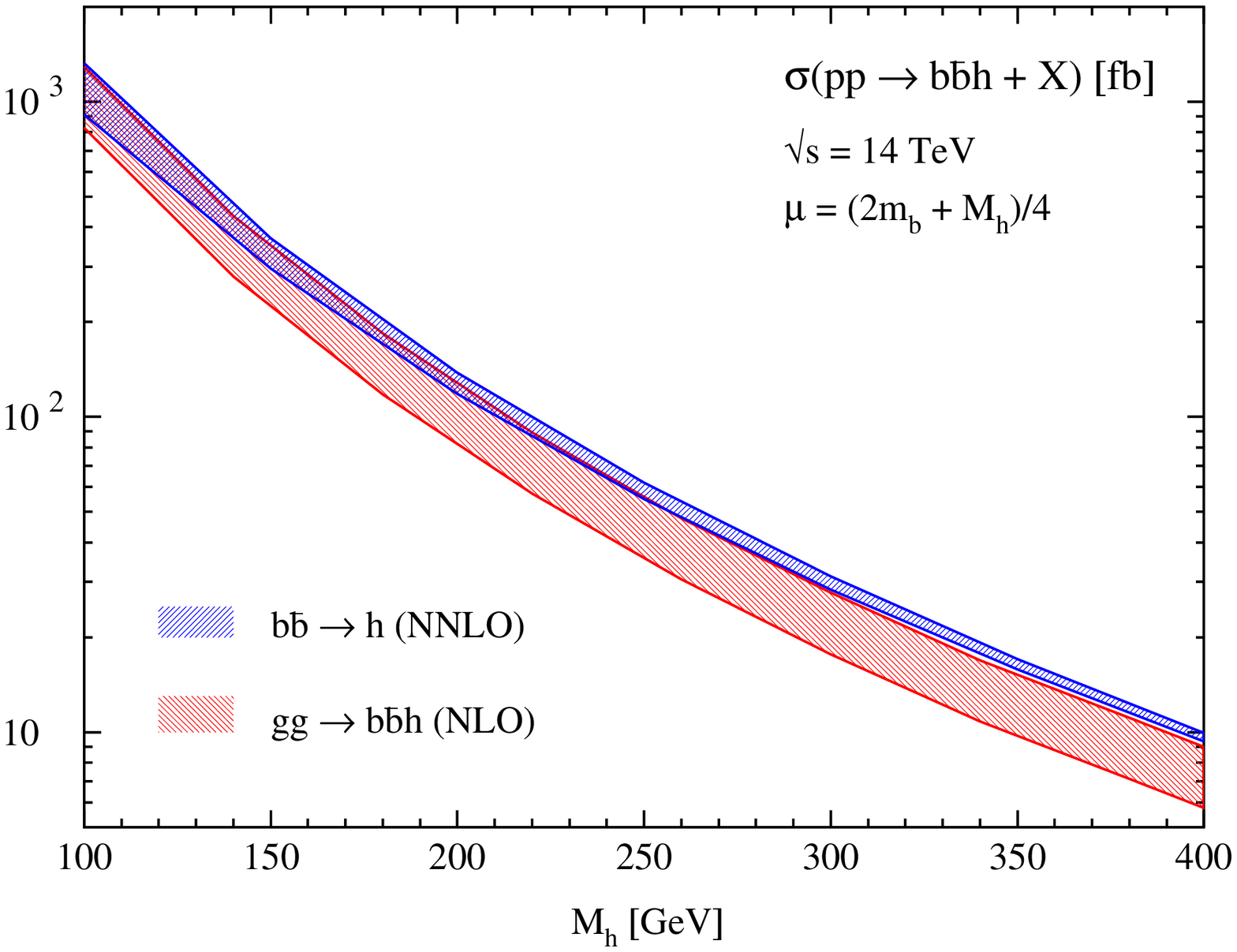}
\caption[]{Total cross sections for $p{\overline p} (pp) \rightarrow b
{\overline b} h+X$ at the Tevatron and the LHC as a function of the
Higgs mass $M_h$ with no $b$ jet identified in the final state.  The
error bands correspond to varying the scale from $\mu_R=\mu_F=(2m_b+M_h)/8$ to
$\mu_R=\mu_F=(2m_b+M_h)/2$.  The NNLO curves are from Ref.~\cite{hk}.}
\label{fg:0b_sigma}
\end{center}
\end{figure}

An alternative calculation is based on the processes $gg\to b\bar b h$
and $q\bar q\to b\bar b h$ (four-flavor-number scheme), which has been
calculated at NLO~\cite{dks,dks2,djrw2}. Despite the presence of the
logarithms $L_b$ in the calculation based on $gg\to b\bar bh$, which
are not resummed, it yields a reliable inclusive cross section, as
evidenced by Fig.~\ref{fg:0b_sigma}. A sizeable uncertainty due to the
renormalization and factorization scale dependence remains which might
reflect that the logarithms $L_b$ are not resummed in this approach,
so that the perturbative convergence is worse than in the
corresponding case of $t\bar th$ production~\cite{tth}.
In the Standard Model,
the closed top-quark loops have been included
in the four-flavor-number calculation and
reduce the inclusive NLO total
cross section for $p p (p{\overline p})
\rightarrow b {\overline b} h$
 by $\sim -4\%$ at the Tevatron  and  $\sim -9\%$ at the LHC for
$M_h=120$~GeV \cite{dks2,djrw2}.
In the MSSM, the closed top quark loops are
negligible for large
$\tan\beta$~\cite{dks,djrw}.

The NLO four-flavor-number scheme calculation is compared with the
NNLO calculation of $b\bar b\to h$ (five-flavor-number scheme) in
Fig.~\ref{fg:0b_sigma}.  The two calculations agree within their
respective scale uncertainties for small Higgs masses, while for large
Higgs masses the five-flavor-number scheme tends to yield larger cross
sections. Note that closed top-quark loops have not been included in
the NNLO calculation of $b\bar b\to h$~\cite{hk}.  

To all orders in perturbation theory the four- and five-flavor number
schemes are identical, but the way of ordering the perturbative
expansion is different and the results do not match exactly at finite
order. The quality of the approximations in the two calculational
schemes is difficult to quantify, and the residual uncertainty of the
predictions may not be fully reflected by the scale variation
displayed in Fig.~\ref{fg:0b_sigma}.

\section{Conclusions}

We investigated $b\bar bh$ production at the Tevatron and the LHC,
which is an important discovery channel for Higgs bosons at large
values of $\tan\beta$ in the MSSM, where the bottom Yukawa coupling is
strongly enhanced~\cite{cdf,atlascms}.  Results for the cross sections
with two tagged $b$ jets have been presented at NLO including
transverse-momentum and pseudorapidity cuts on the $b$ jets which are
close to the experimental requirements. The NLO corrections modify the
predictions by up to $50\%$ and reduce the theoretical uncertainties
significantly.  For the cases of one and no tagged $b$ jet in the
final state we compared the results in the four- and
five-flavor-number schemes.  Due to the smallness of the $b$ quark
mass, large logarithms $L_b$ might arise from phase space integration
in the four-flavor-number scheme, which are resummed in the
five-flavor-number scheme by the introduction of evolved $b$ parton
densities. The five-flavor-number scheme is based on the approximation
that the outgoing $b$ quarks are at small transverse momentum. Thus the
incoming $b$ partons are given zero transverse momentum at leading
order, and acquire transverse momentum at higher order.  The two
calculational schemes represent different perturbative expansions of
the same physical process, and therefore should agree at sufficiently
high order.  It is satisfying that the NLO (and NNLO) calculations
presented here agree within their uncertainties.  This is a major
advance over several years ago, when comparisons of $b\bar b\to h$ at
NLO and $gg\to b\bar bh$ at LO were hardly encouraging~\cite{rsz,tev}.

\section{Acknowledgement}
We thank the organizers of the 2003 Les Houches workshop for organizing
such a productive and interesting workshop.

\end{document}